\documentclass[copyright]{eptcs}
\providecommand{\event}{DSL 2011} % Name of the event you are submitting to

\title{SAGA: A DSL for Story Management}
\author{Lucas Beyak\qquad\qquad Jacques Carette
\institute{Department of Computing and Software,\\
McMaster University, Hamilton, Ontario, Canada}
\email{lbeyak@gmail.com \qquad\qquad carette@mcmaster.ca}
}
\def\titlerunning{SAGA}
\def\authorrunning{L. Beyak \& J. Carette}

\usepackage{color}      % Use colours for code listings
\usepackage{float}      % Allow "H" parameter for figures
\usepackage{graphicx}   % For pictures
\usepackage{syntax}     % for typesetting BNF
\usepackage{hyperref}
\usepackage{listings}   % Allow listing of source codes
\usepackage{wrapfig}    % take less space for figures

\newcommand{\mylabel}[1]{{\bf #1}: }

\begin{document}
\maketitle

\renewcommand{\topfraction}{0.95}
\setlength{\intextsep}{4pt plus 2pt minus 2pt}

\begin{abstract}
Video game development is currently a very labour-intensive endeavour.
Furthermore it involves multi-disciplinary teams of artistic content creators
and programmers, whose typical working patterns are not easily meshed.  SAGA is
our first effort at augmenting the productivity of such teams.

Already convinced of the benefits of DSLs, we set out to analyze the
domains present in games in order to find out which would be most amenable to
the DSL approach. Based on previous work, we thus sought those sub-parts that
already had a partially established vocabulary and at the same time could be well
modeled using classical computer science structures.  We settled on the 'story'
aspect of video games as the best candidate domain, which can be modeled using
state transition systems.

As we are working with a specific company as the ultimate customer for this
work, an additional requirement was that our DSL should produce code that can
be used within a pre-existing framework.  We developed a full system (SAGA)
comprised of a parser for a human-friendly language for 'story events', an
internal representation of design patterns for implementing object-oriented
state-transitions systems, an instantiator for these patterns for a specific
'story', and three renderers (for C++, C\# and Java) for the instantiated
abstract code.
\end{abstract}

\section{Introduction}\label{sec:intro}

The video game industry has become a mainstream phenomenon. Video games are now
as ubiquitous as movies, books, and other forms of popular culture. The
industry's growth might even be outpacing that of the more classical art forms
\cite{growth1,growth2}.  But the development of
video games is currently very labour intensive~\cite{fund,RollingsMorris}.
Some game development companies in our area are actively engaging us to
find better methods for creating interactive digital media products.

To better understand the problem, we need to look into the organizational
structure of medium to large-sized game development studios. At a high level,
there are various common departments that are present in a development studio:
art, music, programming, game design, and management
\cite{game}. These broad departments might be partitioned into smaller sections
depending on the size of the studio. There might be dedicated staff for world
and level design, music and sound effects, user interface design, or story
design to name a few. These employees are domain experts in their respective
fields, but they may know very little of how the other departments operate.
More specifically, neither artists nor musicians, or even level designers,
really need to be programmers.  

However, currently a lot of these tasks still require intense participation
from the development staff.  In other words, the various artistic designers
must communicate their ideas and intent to the programmers.  Not only can this
take much time and effort, it can also easily introduce misunderstandings.
This inefficient and failure-prone communication channel can really 
hamper team productivity.

It would be much simpler, and indeed more efficient, if the designers could
``code'' their ideas in a way that is both natural for them, but also
usable for the production of the game software.  For \emph{some} of the 
domains involved, a domain specific language (DSL) could be a significant part
of the solution.

We believe that some of our requirements (see the next section for the
details) are somewhat different than the
typical problems addressed by DSLs.  Although there are domain-specific
languages that are used by non-programmers to achieve their aims
(Matlab, Verilog, HTML, Mathematica, Maple, LaTeX, Excel, etc), every
single one of the above eventually needs to be \emph{programmed} to 
achieve \emph{large scale} solutions.  We definitely want to avoid this.
Here is an overview of our requirements.
\begin{itemize}
\item  \mylabel{productive} To increase the productivity of domain
experts.
\item \mylabel{comfortable} Provide a comfortable
interface to a non-programmer domain expert.
\item \mylabel{usable} The information thus gathered must be
(automatically) transformable into a usable format for building a game.
\item \mylabel{integratable} The resulting output (code) can be integrated into
an existing game.
\item \mylabel{low-overhead} The computational overhead (in the running game)
for dealing with this game aspect should be low.
\end{itemize}
\noindent  These requirements are not specific to game programming, but
rather are requirements for there to be a good match between an application
domain and a solution based on an external DSL which will be ``compiled''
to code in a traditional programming language and integrated inside a larger
application.

We view our contributions to be: a comfortable DSL for story designers
(who are not assumed to be programmers) that increases the productivity
of a game design team; the start of an abstract language of object-oriented
designs of state transformers; a nice case study of domain analysis from
a domain (game design) which has so far seen few such studies; a 
straightforward design of an external DSL as well as an embedded DSL
which follows established patterns.

The rest of the paper is structured as follows.  First we explain our
requirements in more detail, which leads into an explanation of which
domain we picked for this first foray into DSLs for games -- story
management.  We then explain the abstract structure of interactive stories in 
Section~\ref{sec:structure}, as this gives the semantics that we must
reflect.  Our results need to be integrated into existing development
processes so in Section~\ref{sec:managing} we give an overview of how story
information is used in game development and in game engines.  Combining these
pieces of information, Section~\ref{sec:modeling} explains how we model
stories.  We then give a high-level overview of the design of our solution,
SAGA (\textbf{S}tory as an \textbf{A}cyclic \textbf{G}raph \textbf{A}ssembly),
and how it meets our requirements.  Section~\ref{sec:syntax} gives an overview
of the human-oriented syntax of our DSL.  Section~\ref{sec:modelcode} further
explains one particular part of our design, which involves modeling
\emph{abstract object-oriented code} in a language-independent manner,
followed in Section~\ref{sec:pretty} with explanations of how to generate
human-readable code in multiple languages from the abstract code.  We then
quickly mention a few minor but interesting implementation details.
We finish by reviewing related work, and then drawing some conclusions from
our efforts.

\section{The Domain}\label{sec:which}

We first expand on our requirements, which gives a vocabulary and context
for explaining our chosen domain (story management) in more detail.  In
the lead-up to the current work, we had explored other domains, and the
interested reader can refer to~\cite{BeyakReport} for the details.  The various
aspects of a game are well explained in~\cite{fund}, and gave us a good
starting list of potential domains.  Although this work started as an
independent exploration of the application of DSLs to games, towards the end of
the project we started working with a game design studio.  We have validated
some of our choices with them (external syntax, internal model), but not our
choice of ``communication protocol'' between our modules and the rest of the
engine.  Unfortunately, those details cannot be discussed here because of a
non-disclosure agreement.  

It is very important to note that we are generally interested in 
graphics-heavy, narrative-driven role playing games (RPGs).  There are many
other styles of games, some of which also involve stories, most notably
``interactive fiction''.  Our work is really focused on improving the
productivity of development teams working on RPGs, and may not apply as
readily to other styles of games.

\subsection{Requirements}\label{sec:req}

The \emph{productivity} requirement should be self-explanatory: changing
development methodology should result in a measurable productivity gain
at the level of a \emph{team}, even after including the costs of
additional training and tool development.

The \emph{comfort} requirement states that non-programmer domain experts should
be able to comfortably use the tool that we produce. This stems from the
fact that for \emph{large} video games produced by medium and large-sized
studios, the development team includes \emph{experts} in various domains
(visual arts, writers, musicians, etc) who are not programmers, although
they have skills and talents critical to the development of large-scale games.
Their productivity is seriously hampered if they cannot work with tools
that are familiar to them, and whose basic vocabulary is not the vocabulary
of their domain expertise.  Our discussion with our industry partner has made 
it quite clear that this is where they currently see the largest potential
productivity gains, as this is where they see the most ``friction'' that
delays their development schedules.

But the work of such experts still needs to be \emph{useful} in
creating a game, or put another way, it should be \emph{usable} by the
game programmers for the production of the final executable.  To reduce the
overhead of communication between domain experts and programmers, the
semantics of the language used by the domain experts should be sufficiently
formal as to be directly interpretable.  A further optimization then
consists of compiling this language.

The \emph{integratable} requirement comes from our industry partner:  while
they do desire productivity improvements, they 
have made an already large investment in their current engine and are not
interested in any methodology that would require them to start from scratch.
Thus we need to be able to isolate some components that are segregated from the
rest.

The \emph{low-overhead} requirement (that the CPU time required in the
running game for this aspect of a game should be low) is perhaps the least
obvious.  In fact, based on some of our previous
work~\cite{CaretteKiselyov11,CaElSm11}, it might have been expected that we
would wish to work on (parts of) the physics engine or the graphics rendering
pipeline as a target domain. These domains are fully within the expertize of
well-trained software engineers, and are in fact areas where obtaining
measurable productivity increases would be extremely challenging, albeit the
subject of much ongoing research.  This requirement is in many ways a
\emph{derived requirement} from the `comfort' and `integratable' requirements.
We target non-programmer domain experts, and we want to find components that we
can essentially replace as a whole.

\subsection{Games and Stories}

Story management is not the only part of a game where a DSL could be
fruitfully employed.  We seriously considered three domains which seemed
to be a good fit: character status representation, real-time strategy game
rules, and story management.  We chose story management as we were more
easily able to find reasonable (formal) models in the literature.  This was
a serendipitous choice, as it was also our industrial partner's preferred
starting point.

The story aspect of a game is an important contributing factor to
player immersion \cite{fund}.  Immersion is a very powerful tool to engage
the player with the game world.  Most current games have a linear story,
which do not offer the player much agency during their gameplay.
Agency is the idea that the player has some control over their
surroundings and that their actions have an effect on the game world
\cite{agency}. A player who feels like they are actually interacting with a
game's environment and are able to change it will become immersed more easily.
Agency and immersion in a game will lead to a more involved player experience,
and usually to more enjoyment.

Like in the movies, strong character development or an
engaging storyline can create a very powerful experience for the viewer.
By leveraging the interactivity aspect of a video game, the overall
player experience can be even more immersive.  But if there are many choices in
the storyline, this requires additional assets (cut scenes, dialog, etc)
to be produced which may never be seen by players -- a cost few studios
are willing to bear.

Myst~\cite{myst} is a classic example of a game that makes use of a story that
can be changed by the player, although only at the end of the game.  The
player can make one of four choices, and based on those decisions,
different ending cinematics and dialogue are introduced. Myst also has an
``open-world'' style of gameplay, where the player is free to roam throughout
the game world as they wish, which again creates immersion as the player is
never presented with a ``you can't do that now'' moment. A moment like this can
quickly jar the player out of deep immersion and cause great annoyance
\cite{fund}.

As further examples, Heavy Rain and the Mass
Effect series (Mass Effect 1 and Mass Effect 2), not only let the
player make decisions in the game, but these decisions have important
consequences on the game's story. In Heavy Rain there are four main
characters, but depending on the path that the player takes, some main
characters can be killed and will not return to the game for the remainder
of play \cite{rain}. An interesting aspect of the Mass Effect series is that
the save data from Mass Effect 1 can be used to start a game of the sequel
\cite{mass}. This keeps the story consistent based on the decisions that were
made by the player during the play of Mass Effect 1.

\section{The Structure of Stories}\label{sec:structure}

\begin{wrapfigure}{l}{4cm}
  \centering
    \includegraphics{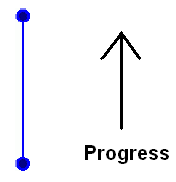}
  \caption{A linear story}
  \label{fig:lineararrow}
\end{wrapfigure}

The story of a video game can be structured in the same ways that a story is
constructed in movies or books. The main difference is that instead of
passively watching characters make decisions, the player
is the one making decisions for the characters.  Note that the main source
material for this section is \cite{scriptP1}, and that all figures originated
from \cite{scriptP1} as well.  This section contains no new contribution,
but understanding the structure of stories is crucial for the development
of a formal model of stories.

\subsection{Linear}

The simplest structure for a story is \emph{linear}: no real choices will be
offered to the player to direct the story.  The player will be explicitly
directed to make ``choices'' with no alternatives.  These linear stories
(see Figure~\ref{fig:lineararrow}) are quite
predictable, which may lead to a low replay value since the player already
knows exactly what is going to happen the second time through. On the plus
side, linear stories are easier to design than their non-linear counterparts.
Developers knows that all the content that
they create will be seen every time the game is played, and so
no effort is wasted on unseen content.

\subsection{Non-linear}

All other story structures are \emph{non-linear}.  At certain points
in the game the player will be presented with a real choice of what to do next,
and the story component of the game will change accordingly.  This is very
interesting as the player now feels like they can affect their game
environment (agency, as introduced earlier).  There are several kinds of 
non-linear stories.

\subsubsection{Branching}
\begin{wrapfigure}{l}{10.2cm}
  \centering
    \includegraphics{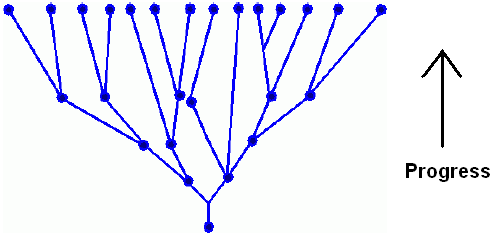}
  \caption{A branching story}
  \label{fig:branchingarrow}
\end{wrapfigure}

A branching story structure (see Figure~\ref{fig:branchingarrow}) is the
standard interpretation of a non-linear story. Each player choice leads
to a completely new part of the story.  While very rewarding from the
player's perspective, this can be rather difficult to implement (in an
RPG with cinematics, etc), due to the amount of content which
must be produced by the game developers. In addition, since
the player actually goes through a single path through the game,
much of the content created for the game goes unseen.  This is not very 
cost effective.  

\begin{wrapfigure}{r}{5cm}
  \centering
    \includegraphics{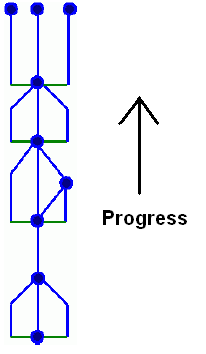}
  \caption{A foldback story}
  \label{fig:foldback}
\end{wrapfigure}
To be more precise, it makes it impractical
to do this in a game with human-created content -- if content was
computationally generated, a branching story may not be such an issue.
Unfortunately, we do not yet know how to generate \emph{immersive}, story-driven
content computationally.

\subsubsection{Parallel Paths (Foldback)}

The parallel paths structure (see Figure~\ref{fig:foldback}) is a compromise
between a linear and a branching structure. The player is still given choices
that affect the story of the game, but at a certain point the story paths
converge, fold back on themselves, to a central point. This central point will
always be reached by the player regardless of what path they choose to take.
After this central point, the player will be given choices again that branch
out for a short time before once again folding back.

This structure gives a good illusion of choice to the player. The player does
have choices that matter but not all the time. Only after playing the game
another time is it possible for a player to realize that there are some
inevitable events that cannot be altered.  This structure offers freedom to the
player while not containing a prohibitive amount of story content to create,
and is the standard story structure in current games~\cite{fund}.

\subsubsection{Threaded}

\begin{wrapfigure}{l}{7cm}
  \centering
    \includegraphics[scale=0.75]{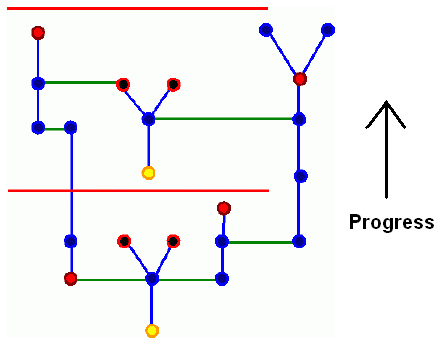}
  \caption{A threaded story}
  \label{fig:threaded}
\end{wrapfigure}

The threaded structure involves multiple independent paths that develop on
their own, regardless of what else is happening in the story of the game. When
the game is nearing the end of the story, these paths usually converge to form
the final events. These are called ``threads'', hence the name of the
structure.  This technique is frequently used in books and movies. 

Figure \ref{fig:threaded} below shows a possible story structure
for a game. This graph is 
be an approximation of the first two acts of the game Discworld Noir. The
yellow dots represent events that start an act. An act is just some structure
to partition the main events of the story. The red horizontal lines separate the
two acts shown in the figure. The blue dots are optional story events, while the
red dots are mandatory. Red dots with black inside are story events that will
make the next act become available. At least one of these special red dots must
occur in each act. The green lines show the areas where threads interact with
one another.

\subsubsection{Dynamic Hierarchical}\label{sec:dynhier}
\begin{wrapfigure}{r}{7cm}
  \centering
    \includegraphics[scale=0.9]{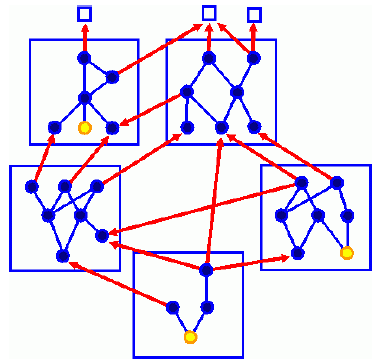}
  \caption{A hierarchical story}
  \label{fig:hierarchical}
\end{wrapfigure}

This structure is called ``dynamic object-oriented'' in~\cite{scriptP2},
but we have chosen to call it ``dynamic hierarchical'' as that is a more
accurate description.  This story structure uses abstraction, by grouping
certain story states into ``sections'', to help manage story complexity.
Figure~\ref{fig:hierarchical} shows a two-level story, with sections
marked by blue boxes.

The general idea at that at each level of the hierarchy, each section is
relatively small and manageable, but still make up a very complex whole
when assembled.  Nesting of sections can theoretically go arbitrarily deep,
but in practice it seems that few levels are needed.
Furthermore, while the story may be large and complex, one can tame the
asset creation problem by re-using locations, and characters, and only
varying aspects which contribute to the narrative.

\section{Stories in Game Development}\label{sec:managing}

We quickly outline how the story aspect of a game enters different parts
of game development: at the specification, design and implementation
stages.

Story progression generally happens in one of three ways: in-game cut-scenes,
fully rendered cinematics, and character dialog.  A story will progress when
some important event or events take place. But once that event has taken
place, this is a permanent change, and the player cannot ``undo'' this
progression.  However, many other things about the setting of the story can
recur: players can come back to a scene they have been in before, talk to
the same people again, even see some cut-scenes again.  But they can never
``unsee'' or ``unhear'' something.

There may be many events that the player can trigger during a game, such as
acquiring some new item or skill, killing creatures, or just exploring an
environment. It is not necessarily the case that all events will actually
affect the state of the game's story. For instance, it might not matter
what kind of sword the player has attained throughout their adventures. It is
important not to confuse the idea of an ``open-world'' style game with a
non-linear storyline. An open-world game means that the player has the freedom
to freely explore a large part of the game world~\cite{open}. Some games
worlds are quite complex, allowing the player to
experience a wide variety of areas in an arbitrary order, which will 
contribute to the player's experience, but nevertheless follow an essentially
linear story -- like the more recent installments of the Zelda series.
Both open-world and non-linear stories will increase the agency of the
player, but are different concepts.

The high-level story structure of a game is primarily a design-level artifact.
From the game code (and data) alone, this structure would be quite difficult
to reverse engineer, at least from the implementations of real-time RPGs
that we know of.  This means that making changes to the story structure,
which happens frequently (or so we are told) is rather expensive.  

The usual representation of progress in a game is done through a (global)
bit vector, where each bit represents an `event' that is important to the
unfolding of the game.  Throughout the
game these event-state variables will be checked to determine if certain
actions can be performed or how to handle various circumstances. There does not
seem to be some overall story ``manager'' being used in current games. Instead
of keeping record of the current story state via a dedicated module, there is
much checking of various boolean state conditions.

Such an implementation hampers reusability and modifiability of the game
engine, at least for aspects relating to narrative and story progress.  It
would be better if there were a module whose
``secret'' \cite{journals/cacm/Parnas72a} is the story
state. This module needs to listen for certain events, and when they occur,
signal to the main game module that the story path has changed. 

It is worthwhile mentioning, however, that processing story events and
transitions is \emph{low-overhead}, in other words requires very little CPU
time (at game run-time).  Since events that are story-relevant are quite rare
(once every few minutes at most), any non-polling implementation will be fine.
The large design-time effort and low run-time CPU requirements for stories make
story management a good target for productivity-enhancing experiments in
alternate development methods.

\section{Modeling Stories}\label{sec:modeling}

The most straightforward (static) model for all types of stories (as defined in
Section~\ref{sec:structure}) is a \emph{directed graph}, with the nodes
representing story states, and the directed edges potential routes from one
state to another.  More precisely, since events cannot ``unhappen'',
story states are sets of events, and transitions happen when specific (sets of)
events have happened. This means that our directed graph is actually
acyclic (i.e. a \emph{directed acyclic graph (DAG)}).  Another way to see
this is that the story graph is actually the graph associated to a poset,
with the elements of the poset given by our states, and the ordering by the
monotone (strict) inclusion relation associate to the transitions.

The dynamic semantics associated to a story are
those of a \emph{finite state machine}.  The transitions are triggered by
certain \emph{events}.  As well, transitions trigger output (the StoryManager
module tells the main module about the new story state), making it into
a (simple) Moore machine.  

In practice, there are two refinements to this model which help.
Some transitions are
triggered only when multiple events occur.  This can be done either by 
having transitions result from a \emph{set} of events, or by having 
nodes that do not correspond to any output, so that these become ``internal''
transitions, which are not externally observable.  Furthermore, to be able to
model dynamic hierarchical stories (see Subsection~\ref{sec:dynhier}), we 
need a hierarchical version of state machines, where states are now sets
of states rather than sets of events.

From this point, one can easily imagine all sorts of extensions to this 
model, such as attaching attributes or computations to nodes or transitions,
allowing random choice of transitions or external triggers for transitions.  We
did not include these extensions as none of them are (currently) present in
single-player RPGs.  In the same vein, while multi-player RPGs are 
very popular, they (currently) do not have a strong ``story'' component,
and thus we consider them out-of-scope.

\section{The Design of SAGA}\label{sec:design}

Our requirements (section~\ref{sec:req}), especially those of
\emph{productivity} and \emph{comfort}, can be well met with an external
DSL which focuses on ``story''.  To support complex stories 
(section~\ref{sec:structure}), it should be possible to express dynamic
hierarchical story scripts in our DSL.  In the next section (\ref{sec:syntax}),
we will give more details of how we have expanded on these requirements.

The \emph{useful} requirement further constrains the DSL: we should be 
able to extract both the underlying DAG of the static story, as well as be
able to extract a state machine model (section~\ref{sec:modeling}) from
the abstract syntax tree of our parsed DSL.  In other words, we need to
be able to interpret a SAGA ``program'' as a specification for a state
transition system.

The \emph{integrable} requirement means that our solution will eventually
need to work well with standard object-oriented languages, primarily
\texttt{C++}, but also \texttt{C\#} and \texttt{Java}.  Rather than
implementing three versions (and likely more in the future), we chose to
implement a code-generator instead.  This choice further frees us to choose any 
language we want for implementing SAGA.

Our experience~\cite{CaretteKiselyov11,CaElSm11,DBLP:journals/jfp/CaretteKS09}
with debugging of code generators has led us to a further design decision:
generated code should not only be human-readable (well-formatted), but
also contain comments aimed to help humans understand the result and
increase traceability.  Our experience in multiple interpretations for
the same embedded language~\cite{DBLP:journals/jfp/CaretteKS09} and
implementing generators for program families~\cite{CaretteKiselyov11,CaElSm11} 
led to another design decision: there should be a single internal model
of the ``generated code'', which can be rendered into our chosen languages.
Furthermore, we should view this internal model also as a language, this time
as an embedded DSL.  It is a sort of ``meta OO'' language specific to the 
\texttt{C}-family of object-oriented languages, and just sufficient for our
purposes.

Summarizing the above, our requirements lead us to implement SAGA as a 
\emph{compiler} from an external DSL focused on ``story'', with the semantics
of specifying certain kinds of state machines to \texttt{C++}, \texttt{C\#}
and \texttt{Java}.  In the next few sections, we will give various
technical details of our implementation.  The reader is always free to
see the complete implementation at~\cite{source}.

\section{Story Syntax}\label{sec:syntax}

\begin{figure}[t]
  \centering
  \fbox{
      \includegraphics[scale=0.8]{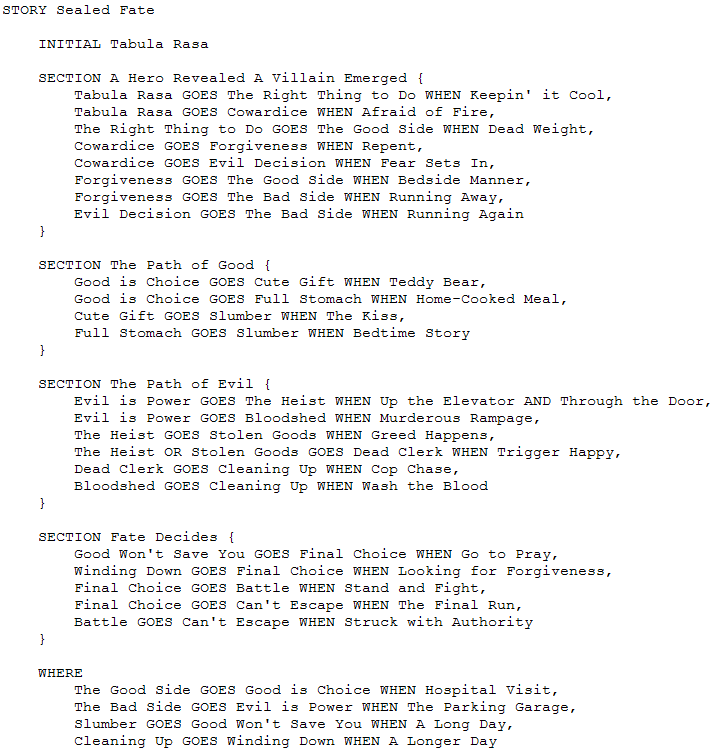}
    }
  \caption{``Sealed Fate'' Story Description}
  \label{fig:storyExample}
\end{figure}

The SAGA syntax needs to be designed from the point of view of a story
designer in order to meet the \emph{comfort} requirement.  We thus
experimented quite a bit on the input language before settling onto what 
we now present.  At first, our language was typical of most programming 
languages with a lot of glyphs as delimiters, including making spaces
significant for lexical analysis.  We eventually settled on a 
language with a few English keywords, minimized glyphs as delimiters, and 
allowing spaces in identifiers.  Figure~\ref{fig:storyExample} gives an
example of a story defined in SAGA.

For example, to have a story description look more like natural language,
we wanted all names of significant items to be phrases.  To be able to 
recognize when these phrases start and end, we used English words as 
delimiters, but capitalized -- rather than quotes or braces.  This makes
these keywords stand out; we chose words that seemed to flow reasonably
well between phrases (like \texttt{GOES} and \texttt{WHEN}) for the principal
delimiters.  The capitalization still makes the keywords stand out from the
labels used to denote names of sections, states, and events.  The whitespace
layout of the language is otherwise completely irrelevant.  The parser also
supports C-style comments in both line and block varieties.

From a story designer's point of view, we can consider a story to be made
up of sections (sometimes called episodes or chapters).  Each section
is made up of a set of distinct states of the story, and of events that may
trigger transitions from one state to another.  At this level, a story
designer only cares about a very restricted set of generally high-level
events that occur in a game, and will want to ignore the majority of 
the irrelevant events of normal gameplay.

The domain expert designing the story will most likely not think about the
story in this way (and even less as a finite state machine!). They will simply
craft the story just as how a writer writes a book. They will think about the
characters, what happens to them, and what the characters will do in order to
overcome their obstacles. The designer will need to be able to identify the
crucial junctures in their story, and separate these into discrete, named
entities.  This is where our design task is at its most subtle -- we want to
minimize implementation issues exposed to the designer, but
still need to be able to use their work directly into the game.

\begin{figure}[th]
  \centering
  	\fbox{
    	\includegraphics{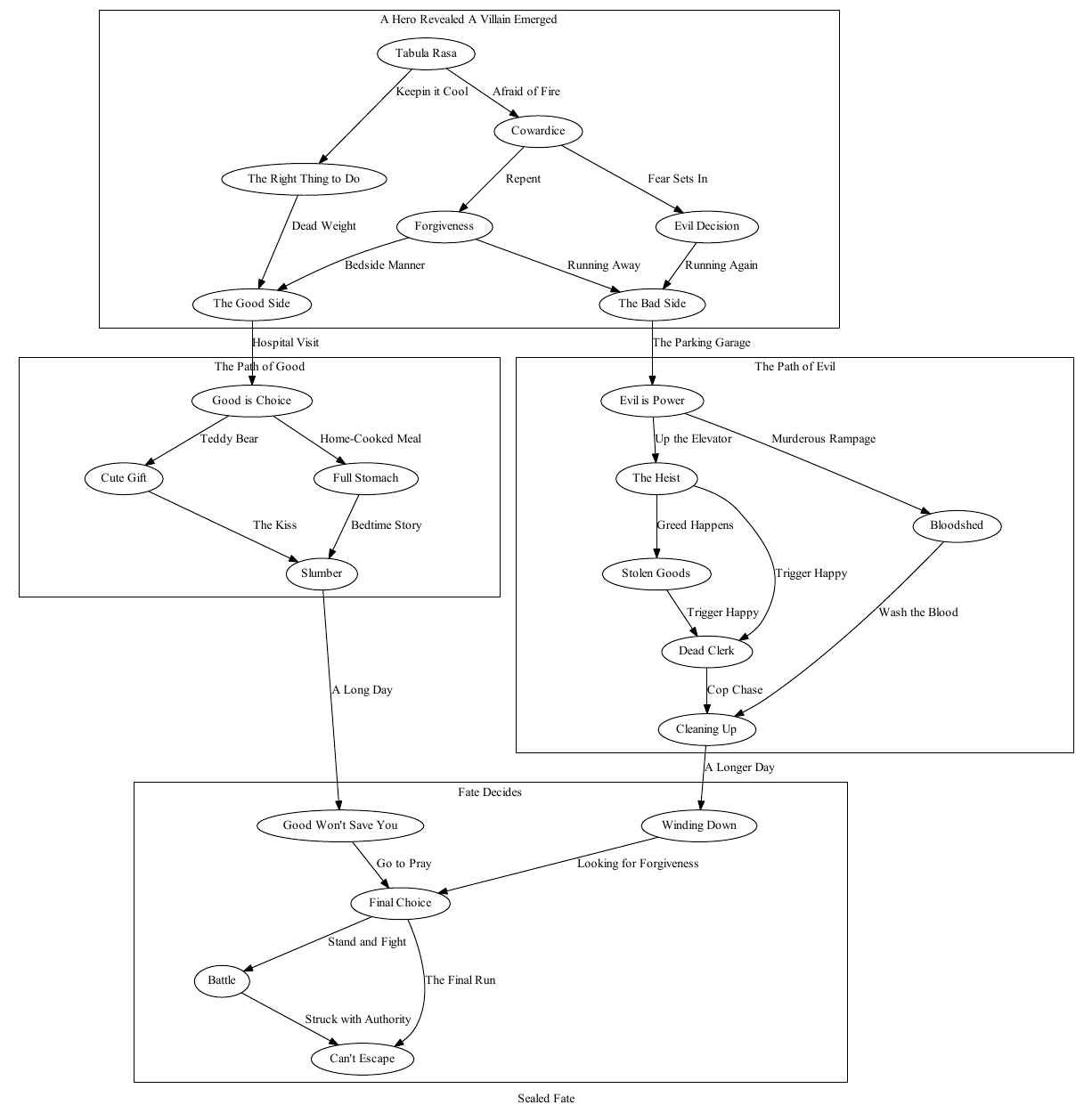}
    }
  \caption{``Sealed Fate'' Story Graph}
  \label{fig:storyGraph}
\end{figure}

Appendix~\ref{lst:ebnf} presents the full syntax of the SAGA DSL.  We will
use the example in Figure~\ref{fig:storyExample} to explain the syntax
(and its semantics) rather than using the EBNF, as that is easier to
comprehend.  The story description in Figure \ref{fig:storyExample} for the
``Sealed Fate'' game story produces a story graph seen in
Figure~\ref{fig:storyGraph}. 

A story must be given a name using \texttt{STORY}, and its starting point 
is given via labeling a node \texttt{INITIAL}\footnote{We assume that our
games all have a singular starting point (although this would be relatively
easy to change).}.  \texttt{SECTION} is used to define (named) sections; the
contents of a section are currently delimited by curly braces, but we intend to
change this.

Sections contain lists of transitions (see the \texttt{<trans>} production
in the grammar).  A transition specifies one or more initial nodes (with
\texttt{OR} as separator for multiple nodes), \texttt{GOES}, a (single)
destination node, \texttt{WHEN}, and a list of events (using \texttt{AND}
as separator).  The interpretation is that when all of the events following
\texttt{WHEN} have occurred, if one is in any of the states indicated
before \texttt{GOES}, the story transitions to the destination state at
that point.  (Uses of \texttt{AND} and \texttt{OR} can be found in the
``The Path of Evil'' section of Figures~\ref{fig:storyExample}
and~\ref{fig:storyGraph}).

The use of \texttt{OR} is not necessary, but it is helpful for 
specifying merge points in a foldback story.  This short form actually
creates separate transitions from each of the listed nodes to the (unique)
destination node.

Transitions between sections are given separately via \texttt{WHERE}.
A \texttt{WHERE} clause is a list of transitions, using the same grammar as
above, but from nodes in separate sections.

At this point, the most obvious extension to a story script is to
include a fuller ``script'' of the story.  From the point of view of story
management, this script has no operational effect, and can be treated as a 
comment (which is why we have omitted this for now).  Of course, from the 
point of view of the overall design of the game, and especially that of the
story designer, this ``script'' is a primary design artifact.

\section{Modeling Code}\label{sec:modelcode}

The SAGA DSL is a purely declarative description of the story domain, 
with an emphasis on transitions between story states.  The DSL itself
provides no indication of how this is to be used in a game.  Thus the 
heart of our compiler is a particular design for implementing a finite
state machine that can be easily embedded into a game.  As we wanted to
target multiple back-end languages, we needed to have an internal model
of an object-oriented meta-language, from which we could produce versions
in \texttt{C++}, \texttt{C\#} and \texttt{Java}.

\lstset{
    language        = Haskell,
    basicstyle      = \footnotesize,
    frame           = single,
    showstringspaces = false
}
\begin{figure}[!bth]
\begin{lstlisting}[label=lst:model]
data AbstractCode = AbsCode Package
data Package = Pack Label [Module]
data Module = Mod {moduleName :: Label, moduleScope :: Scope,
     moduleVars :: [State], moduleFuncs :: [Transformation]}
data Transformation = Transform Label Scope TransType [Parameter] Body
type Body = [Block]
data Block = Block [Statement]
data Statement = AssignState Assignment | DeclState Declaration
     | CondState Conditional | IterState Iteration | JumpState Jump
     | RetState Return | ValState Value | CommentState Comment
data Declaration = VarDec Label StateType
     | ListDec Label StateType Int
     | ListDecLiterals Label Label StateType [Literal]
     | VarDecDef Label StateType Value
     | ObjDecDef Label StateType Value
data Comment = Comment Label | CommentDelimit Label Int
data StateType = NodeT | NodeTransT | SectT | SectTransT | StoryT | StoryManT
               | List StateType | Base BaseType deriving Eq
data TransType = TState StateType | Void | Construct Label
data Scope = Private | Public deriving Eq
data Parameter = StateParam Label StateType
               | FuncParam Label TransType [Parameter]
data State = State Label Scope StateType
\end{lstlisting}
\caption{Definition of Code Model}
\end{figure}

Figure~\ref{lst:model} shows part of our definition of an ``abstract model''
for an object-oriented language.  The largest code unit is a (named)
package, made up a list of modules (aka classes), which themselves
declare a set of state variables as well as a set of functions (aka
methods).  These functions (labeled \texttt{Transformation} in
Figure~\ref{lst:model}) have a signature as well as a body, which is a list
of blocks. Blocks themselves are lists of statements, and this pattern continues.  
These definitions are, by and large, quite standard.
We elide the actual expression language, as it contains nothing surprising.

There are a few things worth pointing out in this model.  For example a
function body is not just a list of statements, but rather a list of blocks. 
This is used for two purposes:
while blocks have no actual semantic meaning, the intent is that it is 
a ``new'' block of code, and as such the renderer can choose to insert
whitespace around a block; secondly, one kind of \texttt{Statement} is a
\texttt{Comment}, which the generator can judiciously insert.  Another is
that we do not enforce that all declarations be done at the top, which 
is not supported by all languages -- this can be dealt with by the renderers.
We also have explicit types for the classes which correspond to nodes, 
sections, etc in our state machine implementation.  This is convenient,
but it is not clear whether this convenience is worth it -- it is anti-modular,
as it ties the renderer to our application.

While the SAGA language is an external DSL, our AbstractCode representation can
be thought of as an internal DSL.  We provide a series of combinators for
generating ``abstract code'', made to resemble a typical OO language.
For example, we have
\begin{lstlisting}
($<) :: Label -> Value -> Value
infixl 5 $<
l $< v = (Var l) `less` v

($.) :: Value -> Function -> Value
infixl 5 $.
v $. f = ObjAccess v f

infixr 5 $=.
($=.) :: Value -> Label -> Statement
($=.) a@(Var _) b = assign a (Var b)
($=.) a@(ObjVar _ _) b = assign a (Var b)
($=.) _ _ = error "$=. operator is only for assigning to ObjVar or Var values"
\end{lstlisting}
\noindent for the binary operation $<$, method access and state variable
assignment\footnote{The error here indicates that we ought to refactor
our values into l-values and r-values} respectively.

We have chosen not to expose this embedded DSL yet, as it is still a bit too
rough at the edges, and is a little too tied to the implementation of
state machines.  The interested reader can see the gory details in the 
\texttt{SAGA.CodeGeneration.AbstractCode} module.  It does allow us to
write ``code'' like that in Figure~\ref{lst:nodeTranClass}, which implements
the NodeTransition class.  The \texttt{Java} code generated by this is 
shown in Appendix~\ref{lst:java}, while the \texttt{C++} is in
Appendix~\ref{lst:cpp}.  This code appears in these appendices exactly as
generated.  The \texttt{Java} code is the verbatim class source file, while the
\texttt{C++} code represents a section of a larger file.  The code has not
been reformatted in any way.

\begin{figure}[t]
\begin{lstlisting}[label=lst:nodeTranClass]
nodeTransModule =
    let mName = "NodeTransition"
        srcNode = "srcNode"
        src = "src"
        dstNode = "dstNode"
        dst = "dst"
        nodeTransEvents = "nodeTransEvents"
        evts = "evts" in
    pubMod mName [
      privVar node            srcNode,
      privVar node            dstNode,
      privVar (List string)   nodeTransEvents
    ] [
      pubFunc (Construct mName) mName [param src node, param dst node, 
                                       param evts (List string)] [
          Block [
            srcNode $= Var src,
            dstNode $= Var dst,
            nodeTransEvents $= Var evts
          ]],
      pubFunc (typ node) "GetSrcNode" [] $ oneLiner $
          returnVar srcNode,
      pubFunc (typ node) getDstNode [] $ oneLiner $
          returnVar dstNode,
      pubFunc (typ $ List string) getNodeTransEvents [] $ oneLiner $
          returnVar nodeTransEvents
    ]
\end{lstlisting}
\caption{Generator for NodeTransition class}
\end{figure}

The creation of this internal AbstractCode DSL makes it
much easier to see and understand the design of the object-oriented code that
will be generated.
Using this EDSL (embedded DSL), we designed a set of abstract OO modules for one possible
implementation of the state transition systems for our story model.

Note that we did not start with an abstract design for this language, and
then wrote pretty-printers for it.  Rather, we started with a generator for
one specific language (\texttt{C++}), and in successive refactoring steps, 
put more information (as well as moving information around) into the AST so
that it was possible to write a renderer for other languages -- first
\texttt{Java}, then \texttt{C\#}.  For example, one can note that in
Figure~\ref{lst:model}, scoping information is attached to methods, which
fits \texttt{Java} very well (see the code in Appendix~\ref{lst:java}), but
not \texttt{C++}, which completely separates this information (see
Appendix~\ref{lst:cpp}).  The main point however is that the needed information
is available and can be easily extracted as needed.

Not all of the StoryManager logic is dependent on the data provided
through a story description.  The specific finite state machine is
story-dependent, but the FSM interpreter is not.  Nevertheless, this too
is implemented via the same mechanism, as it gives us more flexibility
for making changes to various parts of the implementation, for all languages
at once.

\section{Printing Code}\label{sec:pretty}

As was stated in Section~\ref{sec:design}, we made an explicit decision to 
ensure that our output code be human-readable.  Here we give some details on
how we achieved this.  This turns out to be rather easy, and it is surprising
that this is seldom done.

First, it is important to understand and use a good set of pretty-printing
combinators.  For our purposes, \texttt{Text.PrettyPrint.HughesPJ} worked
very well.  While each language has a renderer, there is in fact quite a lot
in common between them (as they all belong to the \texttt{C} family of 
languages), and thus quite a few generic combinators.  Before
giving an example, we need to say a few words about the actual implementation.
We first tried to use a type class for abstracting over renderers, but this
turned out to be quite awkward: the issue here is that the choice of output
is independent of the input (by design), and so we need to pass in this 
choice explicitly, which does not `fit' type classes very well.  Thus we ended
up using a record of renderers (with $49$ entries in our latest version), 
roughly corresponding to the various nodes of the AST itself.  When 
assembling a ``configuration'' (our name for a language-specific renderer),
one can choose to use either a generic or specific renderer for each of
these.

As an example, 
\begin{lstlisting}
moduleDocD :: Config -> FileType -> Label -> Module -> Doc
moduleDocD c f _ (Mod n s vs fs) = vcat [
    scopeDoc s <+> (modDec c) <+> text n <+> lbrace,
    oneTabbed [
        transListDoc f n fs,
        blank,
        stateListDoc vs],
    rbrace]
    where scopeDoc     = scopeDocC c
          stateListDoc = stateListDocC c
          transListDoc = transListDocC c
\end{lstlisting}
\noindent is a generic renderer for a ``module'' (i.e. a class).  We can
see the uses of tabs and blank lines to get a pleasing layout.  We can also
see explicit dispatch through the configuration (the $3$ calls in the
\texttt{where} clause).

As mentioned previously, to be able to insert comments in the appropriate
places in the code, we needed to be able to identify
\emph{conceptual blocks} in the code.  This is where the \texttt{Block} concept
comes from, and is used to separate different conceptual fragments of the code.
These are used by the renderer(s) to produce appropriate whitespace and 
comments.  Forcing ourselves to add such annotations required us to make
our code model even more modular, which was definitely worthwhile.
For example, Figure~\ref{lst:csharp:fate} shows an extract of the \texttt{C\#}
main class, where the actual state machine nodes for the ``Fate Decides''
segment of the story are created and inserted into a story section.  One
can also see that, for increased traceability, the internal names of the
nodes in the code are not ``obscure'', but derived from the phrases given
as the readable node names.  This greatly aids debugging.
\lstset{language=[Sharp]C}
\begin{figure}[th]
\begin{lstlisting}[label=lst:csharp:fate]
    // "Fate Decides"
    Node node__Good_Won_t_Save_You = new Node("Good Won't Save You");
    Node node__Winding_Down = new Node("Winding Down");
    Node node__Final_Choice = new Node("Final Choice");
    Node node__Battle = new Node("Battle");
    Node node__Can_t_Escape = new Node("Can't Escape");
    
    List<Node> nodes__Fate_Decides = new List<Node>(5);
    nodes__Fate_Decides.Insert(0, node__Good_Won_t_Save_You);
    nodes__Fate_Decides.Insert(1, node__Winding_Down);
    nodes__Fate_Decides.Insert(2, node__Final_Choice);
    nodes__Fate_Decides.Insert(3, node__Battle);
    nodes__Fate_Decides.Insert(4, node__Can_t_Escape);
    // End Nodes --------------------------------------------------------------
\end{lstlisting}
\end{figure}

As we mentioned in the previous section, modularization is done differently
in \texttt{C++}: rather than a file for each module, there are just two files,
one header for all the declarations, and an implementation file.  The
\texttt{C++} renderer does this via two passes on the AST, once to render
the header (and ignores everything else), and once to render the code (which
ignores scoping annotations, etc).

\section{Further Implementation Details}\label{sec:implementation}

Finally, it is frequently useful to get a more visual representation of
our work.  Thus we produce \texttt{dot} output representing the 
Story graph (as seen in Figure~\ref{fig:storyGraph}).

For testing purposes, we have implemented ``dummy games''
to serve as drivers for our generated code.  Through
menu selections, we can walk through the story graph manually to make
sure that our generated code works as intended.

\section{Related Work}\label{sec:related}

Regarding the general theory of DSLs, we tried hard to follow conventional
wisdom that is well-explained in the standard survey~\cite{DSLSurvey}, as well
as in the recent book by Fowler~\cite{mfbook}.  We were
also influenced by the Unix way of working with embedded languages, which is
described in Chapter 8 of Raymond's book~\cite{Raymond:2003:AUP:829549},
as well as some of our experience (already cited).

Our general model for stories has multiple precedents.  For example%
~\cite{Brom:2007:SME:1777851.1777856,DBLP:conf/aiide/BalasBAG08} 
use Petri Nets as a formalism for dealing with complex branching stories.
Whether Petri Nets or different kinds of state machines are best-suited 
for story management still seems to be worth investigating.  A simpler DAG,
much like ours, is used by Nelson and Mateas~\cite{SBDM} to encode a story
plot -- but their aim is to optimize player enjoyment by searching through
the plot graph to give players ``hints'' as to what to do next.
Similar ideas are used in~\cite{DBLP:conf/aiide/SullivanCM08} to ``shape
the global experience so that it satisfies the author's expressive goals
without decreasing a player's interactive agency''.  In none of these
cases do the authors worry about the ease of story creation by domain
experts, which ultimately is our main concern.  Since our goals and the
goals of the authors cited above differ significantly, it is very hard to
judge if any of these models are better or worse than another: they may all
turn out to be extremely well-suited to the given application.

Several referees asked about the links with ``interactive fiction'' (IF).
It is true that stories in IF can be very complex, possibly even more 
complex than those we presented here. In particular, Inform $7$%
\cite{Inform7} is a leading language for writing IF.  Even more so than us, 
Inform $7$ is a DSL using a natural language-like syntax.  However, in IF,
the game \emph{is} the story.  Interactive Fiction, unlike graphical RPGs,
are often written by a single author, rather than requiring a full design
studio.  Our motivation stems not from ``story management'', but rather 
from enhancing the productivity of a large, multi-disciplinary development team.
Story design in this setting is around \%5 of the development effort, but
drives almost \%100 of the game development (for narrative-driven RPGs).
As we said in the introduction, it is the component that introduces the
most ``friction'' in the development process, not the most effort.
SAGA is not designed for writing interactive fiction, it is designed to 
allow the designer of the story aspect of a large RPG to work productively,
while seamlessly integrating their work with the rest of the code base.
While we can probably learn a few tricks from DSLs designed for IF, more
productive for our goals will be to augment how much of an RPG-oriented
``story script'' we can productively handle.  It is perhaps worthwhile noting
that IF engines are all interpreters~\cite{if-interp}, while we are aiming
for more compilation.

As far as the use of DSLs in games, 
Tim Sweeney~\cite{DBLP:conf/popl/Sweeney06} clearly documents that there are
different kinds of tasks in a game, which ultimately (in his opinion) will
require different languages to express properly.  There seems to be a
paucity of academic literature on DSLs (and related techniques) applied
to game development.  We did find a language for board game description%
~\cite{Romein97anapplication}, but like us, Furtado, Santos and
Ramhalho~\cite{DBLP:conf/splc/FurtadoSR10} bemoan this fact while documenting
their work on domain analysis for digital games production using product
lines.

\section{Conclusion}\label{sec:conclusion}

Video games are very complex, and have many different aspects (domains) that
can be considered for improvement. In some ways, video game development is
seriously lacking in the application of software engineering principles when
compared to some other industries related to software development. This is
unfortunate, as the complexity of video games makes them perfectly suited for 
the application of modern software engineering principles.

We have found that it is not prohibitively difficult to create a game-centric
DSL for a well understood domain. Little work has been done in this area
because programmers are worried about learning new languages and techniques,
thereby limiting their work output. However, what we have done is unique
because we are focusing on a DSL to be used by content producers rather than
programmers, who are not familiar with already
existing coding techniques. Since the SAGA language was created specifically
for a domain expert in story design, it should be natural to use and not 
constrain their effectiveness but rather improve it.  Once the initial design
of a \emph{story manager} has been agreed to, we can then completely
remove the programmers from the equation, and put the domain experts
fully in charge. Not only will SAGA improve the effectiveness of these story
domain experts, but will offload some work from the programmers, who already
have much to do.

In other words, what we are trying to achieve is not to remove software
engineers from the equation, but rather to focus their efforts where 
significant \emph{software design} needs to happen.  When their critical
skills are no longer needed, we want to be in a position to allow domain
experts (story authors) to be as productive as possible.  Our concern is 
the productivity of the game authoring team as a whole.

Video game development has started to include a bit more variety in the usage
of programming languages. The standard \texttt{C++} is no longer the sole
language used, even though it is still the most popular by a large margin.
Wishing to be ready for future diversity of languages is in small part why we
chose to also render to \texttt{C\#} and \texttt{Java}.  In fact, these
languages turned out to have much more in common than we expected, creating an
abstraction above them was not very difficult, and provided us with a nice test
for our ideas.

We hope that developing our SAGA system will cultivate an improved future for
the video game industry. SAGA is a step towards a more intelligent production
process for game developers of all types.

\bibliographystyle{eptcs}

\bibliography{saga}

\appendix
% old appendix A
%\section{Main Module Code}\label{sec:maincode}
%
%Here we include the code from the Main module in Listing \ref{lst:main}. This
%shows that SAGA is structured as a compiler with a pipeline of standard
%stages.
%
%\lstset{
%    basicstyle      = \footnotesize,
%    backgroundcolor = \color{white},
%    frame           = single,
%    captionpos      = b,
%    caption         = Main Module Code
%}
%
%\lstinputlisting[label=lst:main]{codes/main.hs}
%
% old appendix B
% \section{Module Dependency Diagram}\label{sec:depdiag}
% 
% Figure \ref{fig:dependencyDiagram} shows the dependencies between the modules
% of SAGA.  Note that this is the static dependency of definitions of functions
% and types -- the previous appendix shows the data flow in SAGA.
% 
% \begin{figure}[ht]
%   \caption{SAGA Module Dependencies}
%   \label{fig:dependencyDiagram}
%   \centering
%     \fbox{
%       \includegraphics[scale=0.2125]{images/dependencyDiagram}
%     }
% \end{figure}
% 

\section{EBNF Grammar of the SAGA DSL}\label{lst:ebnf}
% \lstset{
%     basicstyle      = \footnotesize,
%     backgroundcolor = \color{white},
%     frame           = single,
%     captionpos      = t,
%     caption         = EBNF Grammar of the SAGA DSL
% }

\setlength{\grammarparsep}{1pt plus 1pt minus 1pt}

\small
\begin{grammar}
<story> ::= <opt_white> <story_name> <start_node> <sections> `WHERE'
        <trans_list> <opt_white>

<story_name> ::= `STORY' <white> <label> <white>

<start_node> ::= `INITIAL' <white> <label> <white>

<sections> ::= <section> \{ <white> <section> \} <white>

<section> ::= <section_name> <white> `{' <white> <trans_list> <white> `}'

<section_name> ::= `SECTION' <white> <label>

<trans_list> ::= <trans> \{ <opt_white> `,' <opt_white> <trans> \}

<trans> ::= <pre_nodes> <white> `GOES' <white> <label>
        <white> `WHEN' <white> <events>

<sect_trans_list> ::= [ <trans_list> ]

<pre_nodes> ::= <label> \{ `OR' <label> \}

<events> ::= <label> \{ `AND' <label> \}

<label> ::= <word> \{ <white> <word> \}

<word> ::= <char> \{ <char> \}

<white> ::= <white> { <whitespace_char> }

<whitespace_char> ::= ? any white space character ?

<opt_whitespace> ::= [ <white> ]

<char> ::= ? any visible ASCII printable character ? - <invalid_char>

<invalid_char> ::= `,'
\end{grammar}
\normalsize

%\caption{EBNF Grammar of the SAGA DSL}
%  \lstinputlisting[label=lst:ebnf]{codes/StoryManager.tex}

\newpage
\section{Java Code for NodeTransition Class}\label{lst:java}
\lstset{
    basicstyle      = \footnotesize,
    backgroundcolor = \color{white},
    frame           = none,
    language        = java
}
\begin{lstlisting}
package StoryDSL;

import java.util.Arrays;
import java.util.Vector;

public class NodeTransition {
    public NodeTransition(Node src, Node dst, Vector<String> evts) {
        srcNode = src;
        dstNode = dst;
        nodeTransEvents = evts;
    }
    
    public Node GetSrcNode() {
        return srcNode;
    }
    
    public Node GetDstNode() {
        return dstNode;
    }
    
    public Vector<String> GetNodeTransEvents() {
        return nodeTransEvents;
    }
    
    private Node srcNode;
    private Node dstNode;
    private Vector<String> nodeTransEvents;
}
\end{lstlisting}

\section{C++ Code for NodeTransition Class}\label{lst:cpp}
\lstset{
    basicstyle      = \footnotesize,
    backgroundcolor = \color{white},
    frame           = single,
    caption         = b,
    language        = C++
}

\begin{lstlisting}[caption=Part of the header file generated]
namespace StoryDSL {
    class Node;
    class NodeTransition;
    class Section;
    class SectionTransition;
    class Story;
    class StoryManager;
    
    class NodeTransition {
        public:
            NodeTransition(Node* src, Node* dst, vector<string> evts);
            Node* GetSrcNode();
            Node* GetDstNode();
            vector<string> GetNodeTransEvents();
        
        private:
            Node* srcNode;
            Node* dstNode;
            vector<string> nodeTransEvents;
    };
\end{lstlisting}    

\begin{lstlisting}[caption=Part of the cpp file generated]
NodeTransition::NodeTransition(Node* src, Node* dst, vector<string> evts) {
    srcNode = src;
    dstNode = dst;
    nodeTransEvents = evts;
}

Node* NodeTransition::GetSrcNode() {
    return srcNode;
}

Node* NodeTransition::GetDstNode() {
    return dstNode;
}

vector<string> NodeTransition::GetNodeTransEvents() {
    return nodeTransEvents;
}
\end{lstlisting}
\end{document}